\documentclass[lettersize,journal]{IEEEtran}
\usepackage{cite}
\usepackage{amsmath,amssymb,amsfonts}
\usepackage{algorithmic}
\usepackage{graphicx}
\usepackage{textcomp}
\usepackage{xcolor}

\usepackage{multicol}
\usepackage{multirow}
\usepackage{subcaption}
\usepackage{relsize}
\usepackage{mathtools}
\usepackage[linesnumbered,ruled]{algorithm2e}
\begin{document}

\title{Towards Developing and Analysing Metric-Based Software Defect Severity Prediction Model}
\iffalse
\author{{Umamaheswara Sharma B}\\
{Department of Computer Science and Engineering\\
National Institute of Technology\\
Warangal\\
Telangana\\
India\\
Email: uma.phd@student.nitw.ac.in}\\
\and
{Ravichandra Sadam}\\
{Department of Computer Science and Engineering\\
National Institute of Technology\\
Warangal\\
Telangana\\
India\\
Email: ravic@.nitw.ac.in}}
\fi
\author{Umamaheswara Sharma B and Ravichandra Sadam
\thanks{Umamaheswara Sharma B is a senior research fellow with Department of Computer Science and Engineering, National Institute of Technology, Warangal, Telangana, India e-mail: uma.phd@student.nitw.ac.in.}% <-this % stops a space
\thanks{Ravichandra Sadam is an Associate Professor with Department of Computer Science and Engineering, National Institute of Technology, Warangal, Telangana, India e-mail: ravic@.nitw.ac.in.}}% <-this % stops a space
%\thanks{Manuscript received April 19, 2005; revised August 26, 2015.}}
%\author{Umamaheswara Sharma B}%,~\IEEEmembership{Staff,~IEEE,}% <-this % stops a space
%\thanks{Manuscript received April 19, 2021; revised August 16, 2021.}}

% The paper headers
\markboth{IEEE Transactions on Emerging Topics In Computational Intelligence}
{Shell \MakeLowercase{\textit{et al.}}: A Sample Article Using IEEEtran.cls for IEEE Journals}

%\IEEEpubid{0000--0000/00\$00.00~\copyright~2021 IEEE}
% Remember, if you use this you must call \IEEEpubidadjcol in the second
% column for its text to clear the IEEEpubid mark.

\maketitle

\begin{abstract}
In a critical software system, the testers have to spend an enormous amount of time and effort to maintain the software due to the continuous occurrence of defects. Among such defects, some severe defects may adversely affect the software. To reduce the time and effort of a tester, many machine learning models have been proposed in the literature, which use the documented defect reports to automatically predict the severity of the defective software modules. In contrast to the traditional approaches, in this work we propose a metric-based software defect severity prediction (SDSP) model that uses a self-training semi-supervised learning approach to classify the severity of the defective software modules. The approach is constructed on a mixture of unlabelled and labelled defect severity data. The self-training works on the basis of a decision tree classifier to assign the pseudo-class labels to the unlabelled instances. The predictions are promising since the self-training successfully assigns the suitable class labels to the unlabelled instances.

On the other hand, numerous research studies have covered proposing prediction approaches as well as the methodological aspects of defect severity prediction models, the gap in estimating project attributes from the prediction model remains unresolved. To bridge the gap, we propose five project specific measures such as \textit{the Risk-Factor (RF)}, \textit{the Percent of Saved Budget (PSB), the Loss in the Saved Budget (LSB)}, \textit{the Remaining Service Time (RST)} and \textit{Gratuitous Service Time (GST)} to capture project outcomes from the predictions. Similar to the traditional measures, these measures are also calculated from the observed confusion matrix. These measures are used to analyse the impact that the prediction model has on the software project.
\end{abstract}

\begin{IEEEkeywords}
Software Defect Severity Prediction, Software Quality, Evaluation Measures, Self-Training Semi-Supervised Learning, Oversampling.
\end{IEEEkeywords}

\section{Introduction}
\label{Introduction}
Building highly reliable software is always a challenging task for the software quality assurance team, and that costs more time and manpower \cite{pressman2005software, lyu1996handbook}. In this regard, many organisations are spending an enormous amount of money on their test teams to remove/modify the defective code content before releasing the product. However, many software systems are facing maintenance issues due to the improper development of software modules \cite{pressman2005software}. Of these, some issues may require quick assessment and some may require mandatory assessment with less priority. For this, instead of identifying the severity (priority) of the defective modules manually, automation tools such as software defect severity prediction (SDSP) models have been developed in recent years \cite{tan2020bug, kudjo2020effect, malhotra2013severity}.

The predictive models for the defect severity classification mainly utilise the text records to classify the software modules into respective severity classes \cite{menzies2008automated, lamkanfi2010predicting, zhang2016towards, kumar2021predicting, yang2012empirical}. These models utilise text mining approaches to first extract the features from the documented text and then classify the severity of the defective software modules. However, the literature exhibits a little progress towards providing the solution using the multi-class classification approaches without mining the documented records of the software projects \cite{gomes2019bug}.

As an alternative to proposing the traditional text mining approaches or proposing solutions for the methodological aspects of finding the severity of the defective software module, in this work we propose a classification solution using a self-training semi-supervised learning approach. The primary objective of this work is to classify the software module into five different classes, such as \textit{High Severity, Critical, Major, Non-trivial}, and \textit{Clean} from the mixture of labelled and unlabelled data. In this approach, first the available labelled data is over-sampled using a well-known technique called, the adaptive synthetic sampling (ADASYN) \cite{he2008adasyn}, to enhance the minority classes. After obtaining the balanced training data, the self-training semi-supervised learning model \cite{zou2018unsupervised, zou2019confidence, yu2020fine, oymak2020statistical} is implemented on both labelled and unlabelled data. The self-training is an iterative model, which uses the decision tree model as the base learner to assign the pseudo-labels to the unlabelled instances and, at each iteration, using the pre-defined acceptance threshold, high-confidence instances will be added to the original labelled set. In the end, the generated pseudo-labelled training data is fed to the decision tree classifier to observe the performance on the test dataset.

While most of the literature describes the approaches to the defect severity prediction problem, the gap of estimating the project-specific attributes from the prediction model is still present in the literature. To bridge this gap, to understand how far the prediction results are helpful to the project managers, in this work, we propose five project specific measures, such as, \textit{the Risk-Factor (RF)}, \textit{the Percent of Saved Budget (PSB), the Loss in the Saved Budget (LSB)}, \textit{the Remaining Service Time (RST)}, and \textit{Gratuitous Service Time (GST)}. Similar to the traditional measures, these measures are also calculated from the observed confusion matrix of the prediction model. \textit{The RF} is calculated as the amount of risk in the project as a result of the false negatives. \textit{The PSB} and \textit{LSB} are indicative of the savings and the loss of the original savings in the project, respectively. \textit{The RST and GST} measure the amount of time still required to service the damaged code and unnecessary time spent on the project, respectively. To the best of our knowledge, providing interpretable performance in terms of project attributes is novel in the field of software defect severity prediction. 

For this empirical study, we have evaluated the proposed approach on the four software systems collected from the publicly available AEEEM \cite{d2012evaluating} repository. The experimental evaluations are conducted before and after implementing the self-training model (using the decision tree classifier) to observe the difference in the predictive performances. The comparative analysis is made using both the traditional (such as \textit{Accuracy} and \textit{F-measure}) and proposed measures. The experimental results show that the proposed self-training model successfully assigns the class-labels to the unlabelled instances. On average, the proposed self-training model is showing a reduction in the risk of failure of the system and a reduction in the remaining service time. Hence, as a consequence, the software system accounts for increased budget savings.

This work makes the following novel contributions in the field of software defect severity prediction:
\begin{enumerate}
    \item As an alternative to proposing traditional text-mining approaches for severity prediction, we provide a metric-based solution. From the mixture of labelled and unlabelled data, the self-training semi-supervised classification approach tries to classify the software modules into five different classes, such as \textit{high severity, critical, major, non-trivial}, and \textit{clean}.
    \item To understand how far the prediction results are helpful to the project managers, in this work, we propose five project specific measures, such as, \textit{the risk-factor, the percent of saved budget, the loss in the saved budget, the remaining service time}, and \textit{gratuitous service time}. To the best of our knowledge, proposing such project-specific measures is new to the area of software defect severity prediction.
\end{enumerate}
\textit{\textbf{Paper Organization}}: Section \ref{RelatedWork} presents various text mining approaches for the defect severity prediction task. The detailed architecture of the proposed decision tree based self-training semi-supervised learning model is presented in section \ref{Self-training}. Section \ref{empiSetup} provides details of the utilised datasets and traditional and proposed evaluation measures. The empirical results from the proposed model are discussed in section \ref{results}. The section \ref{threats} provides threats to the validity of the proposed framework. And, Section \ref{conclusion} concludes the work and provides potential research directions for this work.
\section{Related Research}
\label{RelatedWork}
Software defect severity prediction models play an important role in the field of software maintenance because of their nature of identifying the severity of the defective module in quick time. In order to solve this problem, many works have been proposed based on mining the defect reports. This section presents a review of the literature that demonstrates the text mining approaches.

Notably, the first work in this research area by Menzies and Markus proposed a SEVERIS model \cite{menzies2008automated} to automatically assign the severity levels to the defective reports. Their work majorly focuses on extracting the general conclusions about the unstructured defect reports of NASA's five Project and Issue Tracking System (PITS) datasets using text mining and machine learning approaches. This work paves the way for new approaches to defect severity prediction.

The defect severity prediction models are majorly implemented in two stages, such as mining the relevant features from the defect reports and, implementing the classification approach on the observed features. Text mining techniques are used to extract the relevant features from the documented defect reports. For this problem, in \cite{yang2012empirical}, Yeng et al. conducted an empirical analysis using various feature selection schemes. The experimental analysis is conducted on the feature selection schemes such as chi-square, information gain, and correlation coefficient using the multinomial Na\"ive Bayes classification approach. They concluded their work by using such feature selection schemes to obtain better prediction performances.

Machine learning techniques are used to categorise the defective documents into various severity levels. In \cite{tan2020bug}, Tan et al. conducted an experiment to predict the severity of the defective reports in the cross project domain using a multi-nominal logistic regression model. In their approach, first they collected defect data based on the question and answer pairs from Stackoverflow. Then, they combined the collected data with the standard datasets such as Mozilla, Eclipse, and GCC. Later, the newly created new data is used to train the multi-nominal logistic regression model. On comparing their approach with the benchmark machine learning models such as \textit{k}-nearest neighbour, Na\"ive Bayes, and long short-term memory, they concluded that they achieved better prediction performances on the enhanced datasets.

A deep learning-based approach was proposed by Ramay et al. in \cite{ramay2019deep} to predict the severity of the defective reports in a cross-project domain. Their approach first utilises natural language processing methods to preprocess the defect reports. Then, for each defect report, they created an emotion score and constructed a vector. Then, they fed the constructed vector as input to train the deep learning model. On the standard benchmark defect severity datasets, their approach outperformed the benchmark models in terms of \textit{F-Measure} values.

In addition, the works such as \cite{kumar2021deep, kumar2021empirical, kumar2021predicting, zhang2016towards, yang2014towards, lamkanfi2010predicting} also discuss the approaches to defect severity prediction, majorly focusing on the aspects of the text mining approaches.

Our proposed approach for severity prediction is inspired by the work of Thung et al. \cite{thung2015active}. In \cite{thung2015active}, Thung et al. proposed a model called LeDex, an active semi-supervised algorithm which predicts defective modules into three families such as Control \& Data Flow, Structural, and Non-Code. In their approach, they first grouped the set of unlabelled data and then labelled it manually with the group developers. Then, they build a classifier to generate confidence on the unlabelled instances. In the semi-supervised learning phase, they learned the obtained labelled and unlabelled sets using the SVM classifier. For the empirical analysis, they utilised the defect reports from the \textit{Mahout, Lucene}, and \textit{OpenNLP} projects. Their empirical results conclude the success of their approach in classifying the defect reports into the respective defect families.

On the contrary, our work does not depend on manually labelling the text reports. And, our task is to predict the software module into five classes, such as \textit{high severity, critical, major, non-trivial}, and \textit{clean}. Also, in the self-training phase, we adapted the decision tree classifier (from the work \cite{tanha2017semi}) as the base learner to build confidence on the selected pseudo-labels. Above all, we also made a contribution towards providing analysis using the project-specific performance measures.
\section{Architecture of Defect Severity Prediction}
\label{Self-training}
\begin{figure}[t]
\label{Workflow}
\centering
\includegraphics[width=\linewidth]{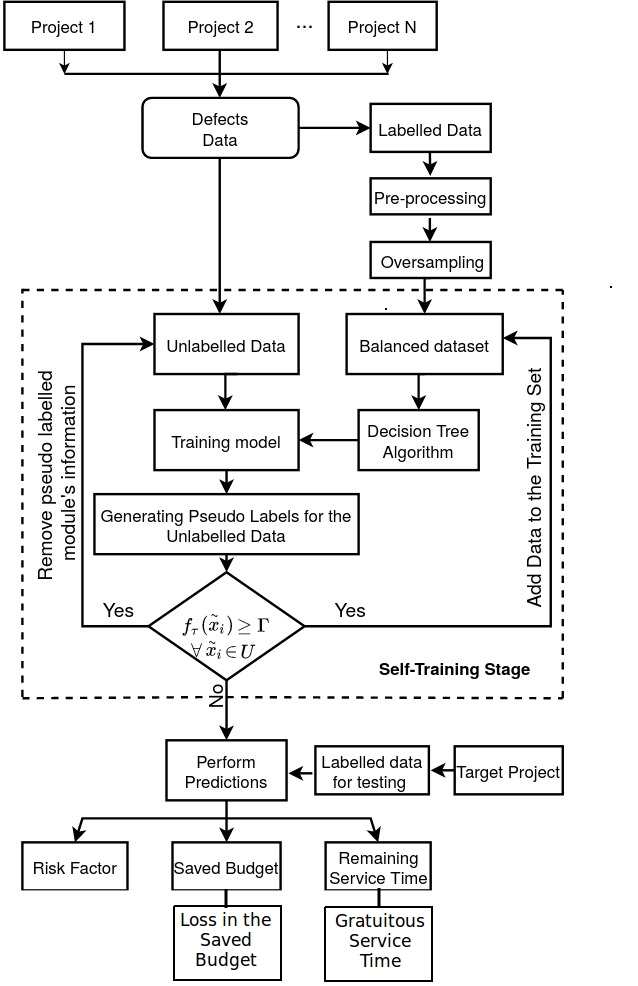}
\caption{An overview of the self-training semi-supervised learning based defect severity prediction}
\end{figure}
In this section, we discuss the details of the proposed architecture of the metric-based software defect severity prediction. The solution approach to this problem is divided into three phases, such as: 1) data preprocessing, 2) oversampling, and 3) self-training semi-supervised learning. The detailed procedure for each phase is given in the following subsections.
\subsection{Preprocessing the Data}
\label{Preprocessing}
For the empirical analysis, this work utilises the defect data of the software projects, collected from the AEEEM benchmark repository \cite{d2012evaluating}. The details of each defect data set are given in the section \ref{studiedDatasets}. Each software project in the AEEEM repository consists of both labelled and unlabelled defect data. Where each \textit{Java class} in the software projects is treated as a software module. And, each software module consists of metrics information along with the number of defects in the severity class. For this learning problem, each software metric is treated as an independent feature, similar to the defect prediction studies \cite{bhutamapuram2021project, Lessmann2008a, Song2011}. 

We have followed the following criteria to identify the labelled and unlabelled data: If the total number of defects is non-zero and the number of defects present in each severity category is zero, then we separate those modules into the unlabelled set. And, the rest of the modules are grouped into a labelled set. To make the problem suitable for multi-class classification, we convert the defect count of each severity class into binary digits. Formally, let $k_{max}$ and $\mathbb{D}_m^{k_{max}}$ represent the maximum possible severity class and the number of defects in that module \textit{m}, respectively. Then the dependent variable (\textit{y}) is then defined for any module \textit{m} as: y = $\mathbb{D}_{m}^{k_{max}}$ = 1, if $\mathbb{D}_{m}^{k_{max}}>$0. And, $\forall k<k_{max}$, y = $\mathbb{D}_{m}^{k}$ = 0. 
\subsection{Oversampling the Minority Classes}
\label{Oversampling}
The major problem that the defect severity models are suffering from is class imbalance distribution \cite{kumar2021predicting}. When compared to the other severity types, it is clear that the majority of software projects rarely encounter highly severe defects \cite{pressman2005software, d2012evaluating}. As a result, the defect data reflects an imbalance among the various severity classes. Consequently, the minority classes may not be predicted successfully because one class dominates the other class, resulting in an underfitting of the model \cite{gomes2019bug}. Underfitting is the most significant impediment to making successful decisions on minority classes \cite{chawla2002smote}.

In this work, we use the widely used Adaptive Synthetic Sampling (ADASYN) approach proposed by Haibo He et al. in \cite{he2008adasyn}, to obtain successful predictions on the rare occurrence of the severity types. The ADASYN is a variant of the Synthetic Minority Oversampling Technique (SMOTE) \cite{chawla2002smote}. The ADASYN works on the basis of weighted distribution on the minority classes according to the level of difficulty in learning whereas, the SMOTE simply generates an equal amount of synthetic examples for each minority class. The ADASYN technique generates synthetic examples of the minority class inversely proportional to the density of the examples in the minority class. That is, this technique generates more synthetic data for the minority classes, which are harder to learn, and less synthetic data for the minority classes, which are easier to learn. The ADASYN improves the learning with respect to the distribution of the data by reducing the bias (which is introduced by the class imbalance) and shifting the class boundary towards the difficult examples.

\subsection{Self-Training Semi-Supervised Learning Approach}
\label{STSSL}
\begin{algorithm}
\SetAlgoLined
\SetKwInput{KwInput}{Input}                % Set the Input
\SetKwInput{KwOutput}{Output}              % set the Output

\KwInput{$\mathcal{S} = \{(x_i,y_i)_{i=1}^{n}\}$: Set of labelled defect severity instances,\\
$\mathcal{U} = \{(x_i)_{i=1}^{u}\}$: Set of unlabelled software modules information,\\
$\mathcal{A}$: Decision-tree classifier,\\
\textit{T}: Maximum number of iterations.
}
\KwOutput{Final Classifier, \textit{f}.}
\SetKwFunction{FMain}{$\mathcal{A}$}
\SetKwProg{Fn}{Function}{:}{\KwRet $f$}
\Fn{\FMain{$\mathcal{S}$}}{
$\mathcal{S} = ADASYN(\mathcal{S})$: // Determine the balanced labelled training set using ADASYN technique\;
$\tau$ = 1\;
\While{($\mathcal{U}!=$Empty)}{ 
	Train the new classifier $f_{\tau} = \mathcal{A}(\mathcal{S})$\;
 	\For{each $ x_i \in \mathcal{U}$}{
 	  Assign the pseudo-label to $x_i$ using the classifier confidence\;
	 }
	 Determine the subset $\mathcal{U}_\tau = (\Tilde{x}_i, \Tilde{y}_i)$ that satisfy $|f_\tau(\Tilde{x}_i)|\geq\Gamma$\;
	 Update $\mathcal{U} = \mathcal{U}-\mathcal{U}_\tau$\;
	 Update $\mathcal{S} = \mathcal{S}\cup\mathcal{U}_\tau$\;
   $\tau$=$\tau$+1\;
   }
Return the final classification model\;
}
Observe the predictions on the target project using \textit{f}.\;
\caption{Self-Training Semi-Supervised Defect Severity Prediction Model}
\label{STSSL-Algorithm}
\end{algorithm}
The SDSP adapts the concept of self-training semi-supervised learning from the works \cite{tanha2017semi, zou2018unsupervised, zou2019confidence, yu2020fine, oymak2020statistical} to predict the severity of the defective software module. The self-training is an iterative method aimed at providing the class labels for the group of unlabelled instances based on the classifier's confidence. That is, the self-training algorithm gradually assigns the most reliable pseudo-labels to the unlabelled instances at each iteration. In self-training, a decision tree model is used as a confidence estimator to build confidence on the generated pseudo-label. A detailed explanation of the self-training procedure is given below.

The algorithmic representation of the proposed model is given in algorithm \ref{STSSL-Algorithm}. The training model takes the labelled defect severity data, which is obtained after oversampling the original imbalanced data. Let us denote the labelled defect training data as $\mathcal{S} = \{x_i, y_i\}_{i=1}^{n}$, where \textit{n} is the size of the labelled instances and, $y_i$ is the discrete representation of the severity class label. Also, let us denote the unlabelled dataset as $\mathcal{U} = \{x_i\}_{i=1}^{u}$, where \textit{u} is the size of the unlabelled instances. Let \textit{f}: $\mathbb{R}^p\rightarrow\mathbb{R}$ be the classification model (in our case, the decision tree classifier), where \textit{p} be the number of features. Let $\hat{y}_f(x)$ be the class label assigned using the confidence estimator \textit{f(x)}. Now, the training algorithm (\textit{f}) will self-train by utilizing the class labels ($\hat{y}_f(x)$). Here, the class labels ($\hat{y}_f(x)$) are also called as the pseudo-labels.\\
As discussed above, self-training is often an iterative approach. In this, the model first utilises the instances where the predictions are more confident, and later it moves to other instances where the predictions are less confident. That is, utilising the more confident pseudo-labels and rejecting the weaker pseudo-labels is the common strategy in the self-training approach. Now, given the classification function (\textit{f}), the function class ($\mathcal{F}$), the acceptance threshold ($1 \geq \Gamma \geq 0$) and, the regularization parameter ($\lambda \rightarrow \infty$), the self-training with the pseudo-labels ($\hat{y}_f(x)$), is targeted to solve the empirical risk minimization problem of the form \cite{zou2018unsupervised, tanha2017semi}:
\begin{multline}
  \hat{f} = \arg \min_{f \in \mathcal{F}} \frac{1}{n} \sum_{i=1}^{n} l(y_i,f(x_i))+\\ \lambda \frac{1}{u}  \sum_{i=n+1}^{n+u}1(f(x_i)\geq\Gamma)l(\hat{y}_f(x_i),f(x_i))
\end{multline}

where the first term in the summation ($\frac{1}{n} \sum_{i=1}^{n} l(y_i,f(x_i))$) refers to the empirical risk of the supervised learning model. And, the second term in the summation ($\frac{1}{u}  \sum_{i=n+1}^{n+u}1(f(x_i)\geq\Gamma)l(\hat{y}_f(x_i),f(x_i))$) refers to the empirical risk of the unsupervised learning model. And, $1(*)$ is the indicator function used to define the happening of the event.

Let us now define the iterative setup of the learning model. Let $\mathcal{A}$ be the decision tree algorithm that takes labelled severity data ($\mathcal{S}$) and builds an initial prediction model, $f_0$. Note that, the initial model is defined on the labelled severity dataset ($\mathcal{S}$) is assumed to be the optimal classifier. Now, with the initial learning model $f_0$ and $\Gamma \in [0,1]$ as the acceptance threshold, then, for a fixed set of iterations \textit{T}, the iterative setup of the learning model operates in the following steps:
\subsubsection{\textbf{Step 1: Generate Pseudo Labels}}
\label{STSSL:Step1}
Determine the subset of severity of the defective software modules $\mathcal{U}_\tau = (\Tilde{x}_i, \Tilde{y}_i)$, $|\mathcal{U}_\tau|<u$, from the unlabelled dataset $\mathcal{U}$ in the current iterate $f_\tau$. Where, $\Tilde{x}_i\in \mathcal{U}$ are the acceptable instances that satisfy $f_\tau(\Tilde{x}_i)\geq\Gamma$. Also, $\Tilde{y}_i$ are the pseudo-labels defined as $\Tilde{y}_i = \Tilde{y}_{f_\tau}(\Tilde{x}_i)$. Here, to eliminate low confidence predictions, $\Gamma\geq0$ is used as the acceptance threshold.
The confidence for the pseudo-labels is derived in the decision tree based on the absolute class frequencies. That is, the estimated confidence value ($f_\tau(\Tilde{x})$) is derived from the majority of the class labels (among all the severity classes) at the leaf node of the trained decision tree. Denoting $K_j$ as the number of number of software modules of the severity type \textit{j} over a total number of instances ($NL$) at the leaf of a decision tree then, the absolute class frequencies of each leaf node of the trained tree is represented as;
\begin{equation}
\label{PseudoLabelEstimator}
   f_\tau(\Tilde{x}) = \frac{K_j}{NL}
\end{equation}

The equation \ref{PseudoLabelEstimator} is also called as the probability estimate and, used to provide confidence on the pseudo-labels for the unlabelled instances. Although, in \cite{tanha2017semi} Tanha et al. suggested different procedures to obtain the improved probability estimates for the unlabelled instances, this work utilises the base probability estimator (the absolute class frequency estimator) as we performed oversampling prior to implementing the self-training procedure.
\subsubsection{\textbf{Step 2: Refine the Classifier}}
\label{STSSL:Step2}
Combine the pseudo-labelled defect severity set $\mathcal{U}_\tau$ and the labelled defect severity dataset $\mathcal{S}$ to obtain a new training set $\mathcal{S}$. Then, build a new decision tree classifier as $f_{\tau+1} = \mathcal{A}(\mathcal{S})$. Now, repeat the step 1 if, $\mathcal{U}$ is not empty.
\section{Empirical Setup}
\label{empiSetup}
Section \ref{studiedDatasets} provides the details of the utilised defect datasets. And, in section \ref{evalMeasures}, the detailed description about the proposed evaluation measures is provided along with the traditional performance measures. Section \ref{Benchmark Classifier} provides the details of the utilised benchmark classifier.
\begin{table*}[]
\centering
\caption{Multi-class confusion matrix}
\label{Multi-ClassConfusionTable}
\begin{tabular}{|l|llllll|}
\hline
 & & & \textbf{Predicted} & & & \\ \hline
\multirow{6}{*}{\textbf{Actual}} & \multicolumn{1}{l|}{} & \multicolumn{1}{l|}{\textbf{High Severity}} & \multicolumn{1}{c|}{\textbf{Critical}} & \multicolumn{1}{l|}{\textbf{Major}} & \multicolumn{1}{l|}{\textbf{Non-trivial}} & \textbf{Clean} \\ \cline{2-7} 
 & \multicolumn{1}{c|}{\textbf{High Severity}} & \multicolumn{1}{c|}{TP1} & \multicolumn{1}{c|}{FN1} & \multicolumn{1}{c|}{FN2} & \multicolumn{1}{c|}{FN3} & FN4 \\ \cline{2-7} 
 & \multicolumn{1}{c|}{\textbf{Critical}} & \multicolumn{1}{c|}{FP1} & \multicolumn{1}{c|}{TP2} & \multicolumn{1}{c|}{FN5} & \multicolumn{1}{c|}{FN6} & FN7 \\ \cline{2-7} 
 & \multicolumn{1}{c|}{\textbf{Major}} & \multicolumn{1}{c|}{FP2} & \multicolumn{1}{c|}{FP3} & \multicolumn{1}{c|}{TP3} & \multicolumn{1}{c|}{FN8} & FN9 \\ \cline{2-7} 
 & \multicolumn{1}{c|}{\textbf{Non-trivial}} & \multicolumn{1}{c|}{FP4} & \multicolumn{1}{c|}{FP5} & \multicolumn{1}{c|}{FP6} & \multicolumn{1}{c|}{TP4} & FN10 \\ \cline{2-7} 
 & \multicolumn{1}{c|}{\textbf{Clean}} & \multicolumn{1}{c|}{FP7} & \multicolumn{1}{c|}{FP8} & \multicolumn{1}{c|}{FP9} & \multicolumn{1}{c|}{FP10} & TN \\ \hline
\end{tabular}
\end{table*}
\subsection{Benchmark Defect Datasets}
\label{studiedDatasets}
\begin{table*}[]
\centering
\caption{The defect severity levels of the modules from the five different projects collected from the AEEEM repository}
\label{DefectsSeverityData}
\resizebox{\textwidth}{!}{%
\begin{tabular}{|cc|c|c|cc|cc|ll|cc|cc|cc|}
\hline
\multicolumn{1}{|c|}{\multirow{2}{*}{}} & \multirow{2}{*}{\textbf{Project}} & \multirow{2}{*}{\textbf{Modules}} & \multirow{2}{*}{\textbf{Total LoC}} & \multicolumn{2}{c|}{\textbf{High Severity}} & \multicolumn{2}{c|}{\textbf{Critical}} & \multicolumn{2}{c|}{\textbf{Major}} & \multicolumn{2}{c|}{\textbf{Non-Trivial}} & \multicolumn{2}{c|}{\textbf{Clean}} & \multicolumn{2}{c|}{\textbf{unlabelled Instances}} \\ \cline{5-16} 
\multicolumn{1}{|c|}{} &  &  &  & \multicolumn{1}{c|}{\textbf{Modules}} & \textbf{Percentage} & \multicolumn{1}{c|}{{\textbf{Modules}}} & \textbf{Percentage} & \multicolumn{1}{l|}{\textbf{Modules}} & \textbf{Percentage} & \multicolumn{1}{c|}{\textbf{Modules}} & \textbf{Percentage} & \multicolumn{1}{c|}{\textbf{Modules}} & \textbf{Percentage} & \multicolumn{1}{c|}{\textbf{Modules}} & \textbf{Percentage} \\ \hline
\multicolumn{1}{|c|}{\textbf{1}} & \textbf{Eclipse} & 997 & 224055 & \multicolumn{1}{c|}{2} & 0.201 & \multicolumn{1}{c|}{10} & 1.003 & \multicolumn{1}{l|}{19} & 1.906 & \multicolumn{1}{c|}{11} & 1.103 & \multicolumn{1}{c|}{791} & 79.338 & \multicolumn{1}{c|}{164} & 16.449 \\ \hline
\multicolumn{1}{|c|}{\textbf{2}} & \textbf{Equinox} & 324 & 39534 & \multicolumn{1}{c|}{0} & 0 & \multicolumn{1}{c|}{1} & 0.309 & \multicolumn{1}{l|}{3} & 0.926 & \multicolumn{1}{c|}{3} & 0.926 & \multicolumn{1}{c|}{195} & 60.185 & \multicolumn{1}{c|}{122} & 37.654 \\ \hline
\multicolumn{1}{|c|}{\textbf{3}} & \textbf{Mylyn} & 1862 & 156102 & \multicolumn{1}{c|}{35} & 1.88 & \multicolumn{1}{c|}{2} & 0.215 & \multicolumn{1}{l|}{4} & 0.215 & \multicolumn{1}{c|}{113} & 6.069 & \multicolumn{1}{c|}{1617} & 86.842 & \multicolumn{1}{c|}{91} & 4.887 \\ \hline
\multicolumn{1}{|c|}{\textbf{4}} & \textbf{PDE} & 1497 & 146952 & \multicolumn{1}{c|}{0} & 0 & \multicolumn{1}{c|}{6} & 3.073 & \multicolumn{1}{l|}{46} & 3.073 & \multicolumn{1}{c|}{7} & 0.468 & \multicolumn{1}{c|}{1288} & 86.039 & \multicolumn{1}{c|}{150} & 10.02 \\ \hline
\multicolumn{2}{|c|}{\textbf{Total}} & 4680 & 566643 & \multicolumn{1}{c|}{37} & 0.791 & \multicolumn{1}{c|}{19} & 0.406 & \multicolumn{1}{l|}{72} & 1.538 & \multicolumn{1}{c|}{134} & 2.863 & \multicolumn{1}{c|}{3891} & 83.141 & \multicolumn{1}{c|}{527} & 11.261 \\ \hline
\end{tabular}%
}
\end{table*}
To evaluate the proposed model, this work utilises the publicly available benchmark defect datasets from the AEEEM repository \cite{d2012evaluating}. From the repository, we have collected severity information about the four software projects, such as \textit{Eclipse, Equinox, PDE}, and \textit{Mylyn}. Each dataset includes the four severity classes of the defective module, as well as information about the \textit{clean} modules. The description of these datasets, such as the number of software modules in each severity category, lines of code (of both labelled and unlabelled modules), and the percent of modules in each severity category, is given in table \ref{DefectsSeverityData}. The table \ref{DefectsSeverityData} also provides information about the number of unlabelled instances which are present in each dataset. A total of 4153 labelled and 527 unlabelled instances were utilised in this experiment. Out of 4153 labelled instances from all the five projects, 37 are the \textit{high severity}, 19 are the \textit{critical}, 72 are \textit{major}, 134 are \textit{non-trivial} defects, and 3891 are the \textit{clean} modules. 
\subsection{Performance Measures}
\label{evalMeasures}
Interpretable performance measures are helpful in understanding the actual behaviour of the learning model. To understand how far the prediction results are helpful to the project managers for the SDSP problem, this work proposes five project specific measures, such as, \textit{the risk-factor (RF)}, \textit{the percent of saved budget (PSB), the loss in the saved budget (LSB)}, \textit{the remaining service time (RST)}, and \textit{gratuitous service time (GST)}. The basis for the proposed measures is to fulfil the primary goal of the SDSP, which is to minimise the total allocated budget by reducing the total time spent on the project \cite{d2012evaluating, bhutamapuram2021project}. The proposed measures are calculated based on the information from the confusion matrix along with the information from the lines of code (LoC). In addition to the proposed measures, the performances are also evaluated using \textit{Accuracy} and \textit{F-Measure}.

The confusion matrix is the basis for all the performance measures. The confusion matrix is constructed based on the actual and predicted labels of the five different classes. The multi-class confusion matrix for this classification task is given in the table \ref{Multi-ClassConfusionTable}. The values of true positives (TP), true negatives (TN), false positives (FP), and false negatives (FN) are calculated for all the five classes. The TP values of any category indicate that the test result correctly predicts the presence of the condition. The TN values of any category indicate that the test result correctly predicts the absence of the condition. The FP values of any category indicate that the test result wrongly predicts the presence of the condition. And, the FN values of any category indicate that the test result wrongly predicts the absence of the condition. The detailed description of the performance measures is given below.
\subsubsection{The \textit{Accuracy}}
\label{Accuracy}
The \textit{accuracy} is calculated as the percentage of the perfect predictions from the total tested instances.
\begin{equation}
    \text{\textit{Accuracy}} = \frac{TP1+TP2+TP3+TP4+TN}{|n_t|}
\end{equation}

where $|n_t|$ denotes the total number of modules in the target project. Since the data is balanced at the preprocessing stage, the \textit{Accuracy} is a well suited measure for this classification task. 
\subsubsection{The F-Measure}
\label{F-Measure}
The \textit{F-Measure} is calculated based on the harmonic mean of precision and recall. This metric indicates how precise (the number of instances correctly classified by the prediction model) and robust (the prediction model does not miss a significant number of instances) the prediction model is. This measure is calculated as:
\begin{equation}
    \text{\textit{F-Measure}} = 2*\frac{Precision*Recall}{Precision+Recall}
\end{equation}
Where, \textit{Precision} is defined as the fraction of the relevant instances (most commonly, defective instances) from the predicted defective instances and, \textit{Recall} is the fraction of the relevant instances from the original defective instances.
\subsubsection{The Risk-Factor}
\label{Risk-Factor}
This measure provides information about the amount of risk present in the project due to the occurrence of false negative instances. Because the severity defines the level of urgency with which the damaged code must be serviced \cite{menzies2008automated}, the misclassification of highly severe (of any category of severity) defective modules into less severe defective modules causes the delay in servicing the damaged code.  Therefore, it may affect the software system at an early stages \cite{lyu1996handbook}. Thus, in this work, to observe the amount of misclassification of highly severe defective modules into less severe defective modules, the \textit{Risk-Factor} is calculated for each category of the severity class. The \textit{Risk-Factor} for each category of severity is calculated as:
\begin{equation}
\label{RF}
    RF(r) = \frac{\sum_s FN_{r,s}*|s-r|}{N_r}, \text{for } \textit{r$<$s}, \forall r,s \in \{Classes\}.
\end{equation}

where, from table \ref{Multi-ClassConfusionTable}, $FN_{r,s}$ represents the number of defective software modules from the $r^{th}$ class are predicted as being from the $s^{th}$ class and, $N_r$ represents the number of defective software modules in $r^{th}$ class. The inequality, $r<s$, ensures the false negative instances from the confusion matrix. 

Since each severity class represents the level of granularity of the defective software module, the hierarchy of each severity class can be represented as an ordinal class \cite{cardoso2011measuring}. In the equation \ref{RF}, when the weight is $|(s-r)|$, wherein \textit{c} is the predicted class value, and \textit{r} is the ground truth of an enumerated ordinal class value, it is equivalent to Mean Absolute Error (MAE) \cite{cardoso2011measuring}. Therefore, it provides the theoretical justification while facilitating the design choice of selecting the appropriate weights (ordinal values) based on the problem.

In our experiments, for a simple case, we define the ordinal values for \textit{high-severity, critical, major, non-trivial}, and \textit{clean} classes as 0.1, 0.2, 0.3, 0.4, and 0.5 respectively. Note that, choosing the ordinal values for the severity class labels is purely a random choice and any research practitioner may replace the above values with the most suitable values in the experiments. Below few examples illustrate the risk factor values in terms of each severity class. For example, if the actual status of the defective software module is \textit{high severity} and if it is classified as \textit{clean} then, the module gets the highest weight of 0.4. This indicates that the defective module is a high-failure prone software module. Also, if the actual status of the defective software module is \textit{high severity} and if it is classified as \textit{critical} then, the module gets the lowest weight of 0.1 because this misclassification may cause less severe damage to the system when compared with the other levels of misclassification. Similarly, there are several cases where the system gets the lowest weight of 0.1. The table \ref{Risk-Factor Values} represents the minimum and maximum possible \textit{Risk-Factor} values for each severity class.

The sum of the \textit{Risk-Factor} values of all the severity classes (from the table \ref{Risk-Factor Values}) represents the system's risk. The last row of the table \ref{Risk-Factor Values} represents the system's \textit{Risk-Factor} obtained by summing \textit{Risk-Factor} values of individual severity classes. Since we have taken the ordinal values 0.1, 0.2, 0.3, 0.4, and 0.5 for the severity classes such as \textit{high-severity, critical, major, non-trivial}, and \textit{clean}, respectively, then from the equation \ref{RF}, the system may have a maximum \textit{Risk-Factor} of 1 and a minimum of 0. In the ideal case, the system with zero \textit{Risk-Factor} indicates that it is in a safe state with no defects (consequently, failures). However, for any machine learning model, it is not always possible to obtain a perfect outcome on the test data \cite{shalev2014understanding}. Thus, minimising the risk is an essential requirement for any prediction model.

In summary, if the prediction model minimises the false negative instances on the test set, then, consequently, it minimises the chances of risk of failure of the software.
\begin{table}[]
\caption{The range of the \textit{Risk-Factor} values.}
\label{Risk-Factor Values}
\centering
\begin{tabular}{|l|c|c|}
\hline
  & \textbf{Severity Class} & \textbf{Range}  \\ \hline
\textbf{1} & \textbf{High Severity}  & $0\leq RF(\text{HS})\leq 0.4$ \\ \hline
\textbf{2} & \textbf{Critical}       & $0\leq RF(\text{Critical})\leq 0.3$  \\ \hline
\textbf{3} & \textbf{Major}       & $0\leq RF(\text{Major})\leq 0.2$  \\ \hline
\textbf{4} & \textbf{Non-Trivial}    & $0\leq RF(\text{Non-Trivial})\leq 0.1$\\ \hline
& \textbf{System's \textit{Risk-Factor}}    & $0\leq RF(\text{Software})\leq 1$  \\ \hline
\end{tabular}
\end{table}

\subsubsection{The Percent of Saved Budget}
\label{Saved Budget}
This measure helps the project managers to estimate the savings in the total allocated budget from the defect severity prediction models. To estimate the savings in the total allocated budget, let us define the percent of true negatives:
\begin{equation}
\label{PTN}
    \text{PTN} = \frac{\text{True Negatives}}{\text{Total Test Instances}} = \frac{|TN|}{|n_t|}
\end{equation}

Note that the true negatives in the confusion matrix indicate the number of \textit{clean} modules which are predicted into their correct class. It is evident that each module in the true negative category can save the amount of inspection needed to be done by the tester \cite{perera2020defect}. Therefore, the tester need not inspect the lines of code of each module in the true negative category.

Now, let us define the \textit{percent of saved budget} measure, which is derived based on the following assumption:\\
\textbf{\textit{Assumption 1:}} \textit{Since the size of any target project is measured through the total lines of code (LoC) \cite{pressman2005software}, for a total budget on the target project, we assign a uniform cost of 1 unit to each LoC.}

Now, the total LoC of the target project = $\sum_{i\in n_t} LoC_i$, where $LoC_i$ is the LoC of the module \textit{i}. Based on the above assumption, the total cost of the target project is derived as $\sum_{i\in n_t} C(LoC_i)$, where, $C(LoC_i)=LoC_i, \forall i\in n_t$, is the cost function. Since true negatives (TN) represent the correct predictions for the \textit{clean} modules, the \textit{percent of saved budget} (PSB) is derived as:
\begin{equation}
\label{PSB}
    \text{PSB} = \frac{\mathlarger{\sum}_{i\in TN} C(LoC_i)}{\mathlarger{\sum}_{i\in n_t} C(LoC_i)}
\end{equation}

Now, the equation \ref{PSB} represents the percent of savings in the allocated budget for the project. The numerator of the equation \ref{PSB} represents \textit{the amount of saved budget} and it is given as:
\begin{equation}
\label{SB}
    \text{Saved Budget} = \mathlarger{\sum}_{i\in TN} C(LoC_i)
\end{equation}

\subsubsection{The Loss in the Saved Budget}
\label{LossSB}
Since the ground-truth value for the modules in $\{\cup_{j=7}^{10} FP_j\}$ is \textit{clean}, then the amount of \textit{loss in the saved budget} (LSB) is derived as: 
\begin{equation}
\label{LSB}
    \text{LSB} = \frac{\mathlarger{\sum}_{i\in \{\cup_{j=7}^{10} FP_j\}} C(LoC_i)}{\mathlarger{\sum}_{i\in n_t} C(LoC_i)}
\end{equation}

In general, the sum of equations \ref{PSB} and \ref{LSB} represents the original budget savings.  The target project will benefit from the machine learning (ML) model if it increases the PSB by simultaneously decreasing the LSB value.

\subsubsection{The Remaining Service Time}
\label{The Remaining Service Time}
This measure helps the project managers to estimate the amount of service required to repair the damaged code in the software project. In contrast to the \textit{percent of saved budget} measure, this measure is defined based on the ratio of the sum of all true positives, false positives, and false negatives over all the tested modules. To define \textit{the remaining service time}, let us define the \textit{percent of non-true negatives (PNTN)} as:
\begin{equation}
\label{PNTN}
    \text{PNTN} = \frac{\text{Non True Negatives}}{\text{Total Test Instances}} = \frac{|n_t|-|TN|}{|n_t|}
\end{equation}

Equation \ref{PNTN} defines the cases where the tester has to spend a quality amount of time on the modules (except true negatives) to identify/remove the defective code from those software modules. Note that, even though the modules in \textit{FN4, FN7, FN9}, and \textit{FN10} indicate the predicted clean modules, after the release of the software, these modules may require inspection because their actual condition is one among the different severity classes. In other words, the tester needs to inspect each line of code of all the modules which are present other than in the true negative category.

Now, the \textit{remaining service time} measure is derived based on the following assumption:\\
\textbf{\textit{Assumption 2}}: \textit{Similar to \textbf{Assumption 1}, for the total service time required on the project, we assign a uniform time of 1 unit to each LoC.}

Based on the above assumption, the time required for the code walk of the modules is derived as $\sum_{i\in n_t} R(LoC_i)$, where $R(LoC_i)=LoC_i, \forall i\in n_t$, is the time function. The modules in $n_t-$TN are the representative of repair/code walk. Now, the \textit{percent of remaining edits (PRE)} is calculated as:
\begin{equation}
\label{PRE}
    \text{PRE} = \frac{\mathlarger{\sum}_{i\in n_t-TN} R(LoC_i)}{\mathlarger{\sum}_{i\in n_t} R(LoC_i)}
\end{equation}

Note that all false negatives are indicative of the ground-truth defective modules. Hence, its service, even after releasing the product, is inevitably required \cite{lyu1996handbook}. Now, the numerator of the equation \ref{PRE} represents the \textit{remaining edits} in the project, which is given as:
\begin{equation}
\label{RemaningEdits}                                  
    \text{\textit{Remaining Edits}} = \mathlarger{\sum}_{i\in n_t-TN} R(LoC_i)
\end{equation}

Assume the test team can modify $\Delta$ LoCs every hour, then the number of project hours required to service the defective modules (\textit{remaining service time} (RST)) is calculated from the numerator of the equation \ref{PRE} as:
\begin{equation}
\label{ProjectHours}
    \text{RST} = \Bigg(\frac{\mathlarger{\sum}_{i\in n_t-TN} R(LoC_i)}{\Delta} \hspace{0.1cm}\Bigg) hours
\end{equation}
\subsubsection{The Gratuitous Service Time}
\label{GratuitousST}
Since the time spent to check for the original status of the modules in $\{\cup_{j=7}^{10} FP_j\}$ is considered unnecessary, then the \textit{gratuitous service time} (GST) for the modules in $\{\cup_{j=7}^{10} FP_j\}$ is derived as:
\begin{equation}
\label{PGE}
    \text{GST} = \Bigg(\frac{\mathlarger{\sum}_{i\in \{\cup_{j=7}^{10} FP_j\}} R(LoC_i)}{\Delta}\Bigg) hours
\end{equation}

Note that the ground truth value of the other false positives such as \textit{FP1, FP2}, $\cdots$, \textit{FP6} is defective. Hence, these modules do not impose an extra burden on the tester.

In general, the difference between the equations \ref{ProjectHours} and \ref{PGE} results in the \textit{original service time}. The target project will benefit from the ML model if it decreases the GST by simultaneously decreasing the RST value.

\subsection{Benchmark Classifier}
\label{Benchmark Classifier}
As discussed in section \ref{STSSL}, the self-training model uses a decision tree classifier as the base learner to build confidence on the assigned pseudo-labels. The details of the decision tree classifier are given below.\\
\textbf{The Decision Tree Classifier}: The self-training utilises the Classification and Regression Trees (CART) \cite{breiman2017classification} model. The CART is used for both regression and classification tasks. The fundamental idea of this technique is to partition the input space into rectangular regions where, for the classification problem, the same class label is given for each region. In the self-training technique, in the process of learning on the pseudo-labelled training data, the tree will split until all the leaves are pure or until all the leaves contain less than 2 samples. 
\section{Results and Discussion}
\label{results}
The empirical evaluation is carried out in the following steps: First, for the comparative analysis, a supervised learning model is built with the original labelled training set. Then the self-training semi-supervised learning is implemented on both the labelled and unlabelled datasets. Later, we compare both the models. This comparative analysis is intended to show the impact of the self-training semi-supervised learning on the unlabelled instances. In the section \ref{Comparative Analysis}, we present the comparative analysis using the proposed performance measures and the traditional measures (such as \textit{Accuracy} and \textit{F-Measure}). The discussion on the research outcomes is presented in the section \ref{Discussion}.
\subsection{Comparative Analysis}
\label{Comparative Analysis}
\subsubsection{Performances in terms of Risk-Factor}

% Please add the following required packages to your document preamble:
% \usepackage{multirow}
\begin{table*}[]
\centering
\caption{The Risk-Factor Values}
\label{HeatMap}
\resizebox{\textwidth}{!}{
\begin{tabular}{cccccccccccc}
\hline
\multirow{2}{*}{\textbf{S.No}} & \multirow{2}{*}{\textbf{Target Project}} & \multicolumn{2}{c}{\textbf{High Severity}} & \multicolumn{2}{c}{\textbf{Critical}} & \multicolumn{2}{c}{\textbf{Major}} & \multicolumn{2}{c}{\textbf{Non-Trivial}} & \multicolumn{2}{c}{\textbf{\begin{tabular}[c]{@{}c@{}}System's \\  Risk-Factor\end{tabular}}} \\ \cline{3-12} 
 &  & \textbf{BST} & \textbf{AST} & \textbf{BST} & \textbf{AST} & \textbf{BST} & \textbf{AST} & \textbf{BST} & \textbf{AST} & \textbf{BST} & \textbf{AST} \\ \hline
\textbf{1} & \textbf{Eclipse} & 0.3 & \textbf{0.15} & 0.19 & 0.22 & 0.1526 & \textbf{0.1421} & 0.0636 & 0.0727 & 0.7062 & \textbf{0.5848} \\ \hline
\textbf{2} & \textbf{Equinox} & 0 & 0 & 0.3 & 0.3 & 0.1667 & 0.1667 & 0.1 & 0.1 & 0.5667 & 0.5667 \\ \hline
\textbf{3} & \textbf{Mylyn} & 0.3486 & 0.36 & 0.1 & 0.1 & 0.2 & \textbf{0.1} & 0.0699 & 0.0814 & 0.7185 & \textbf{0.6414} \\ \hline
\textbf{4} & \textbf{PDE} & 0 & 0 & 0.1833 & 0.25 & 0.1783 & \textbf{0.1696} & 0.0714 & 0.0857 & 0.433 & 0.5053 \\ \hline
\multicolumn{1}{l}{} & \textbf{Average} & 0.1622 & \textbf{0.1275} & 0.19 & 0.22 & 0.1744 & \textbf{0.1446} & 0.0762 & \multicolumn{1}{l}{0.08495} & 0.6061 & \textbf{0.5746} \\ \hline
\end{tabular}
}
\end{table*}

In the table \ref{HeatMap}, we present the \textit{Risk-Factor} values of each target project in terms of each severity class, obtained by using the proposed decision tree-based self-training approach. In table \ref{HeatMap}, the \textit{Risk-Factor} values for the \textit{high severity} class in the target projects \textit{Equinox} and \textit{PDE} are given as 0 because these projects do not have \textit{high severity} modules. 

It is observed from the table \ref{HeatMap} that, with the use of self-training, the risk of failure of the software with respect to the  \textit{High Severity} and \textit{Major} classes is decreased in the target project \textit{Eclipse}. Whereas in the project \textit{Mylyn}, there is a decrement in the risk of failure of the software with respect to the \textit{major} severity class. The project \textit{PDE} records a decrease in the \textit{Risk-Factor} values in terms of Major severity class. For the project \textit{Equinox}, it is observed that, there is no change in the \textit{Risk-Factor} values (in terms of all the severity classes).

When we consider the risk of a system's failure, the projects such as \textit{Eclipse} and \textit{Mylyn} (from table \ref{HeatMap}), benefit from the use of self-training as there is a decrease in the risk with a difference of 0.1214 and 0.0744, respectively. For the project, \textit{Equinox}, the self-training model does not show any difference in the risk of system's failure. And, with the use of self-training, there is an increase in the risk of failure of the system, \textit{PDE}. 

On the other hand, on average, the risk of system failure when the high-severe modules are misclassified into other low classes is reduced by 0.0347 points. Similarly, on an average, the risk of system failure when the major modules are misclassified into other low classes is reduced by 0.0298 points. In contrast, on an average, the risk of system failure when the critical modules are misclassified into other low classes is increased by 0.03 points. Similarly, on an average, the risk of system's failure, when the non-trivial modules are misclassified into the clean class, is increased by 0.0088 points.

On the other hand, on an average, the self-training reduces the system's risk by about 0.0315 (0.6061-0.5746). This indicates the model's ability to successfully reduce false negative predictions. However, on an average, the target projects still have a risk of 0.5746, which is still problematic as it has a 57.46\% chance of system failure. Nullifying the system's risk-factor is the only criteria to avoid the failures in the system. However, achieving a failure-free system is possible only in an ideal case. Hence, any future defect severity prediction studies should be targeted to minimise the risk of system failure.
\subsubsection{Performances in terms of Other Measures}
\begin{table*}[]
\centering
\caption{Performance of the SDSP using the traditional and proposed metrics. Here, BST indicates Before Self-Training and AST indicates After Self-Training}
\label{Performances}
\resizebox{\textwidth}{!}{
\begin{tabular}{cccccccccccccccc}
\hline
\multirow{2}{*}{\textbf{S.No}} & \multirow{2}{*}{\textbf{Target Project}} & \multicolumn{2}{c}{\textbf{\textit{Accuracy}}} & \multicolumn{2}{c}{\textbf{\texttt{F-Measure}}} & \multicolumn{2}{c}{\textbf{PSB}} & \multicolumn{2}{c}{\textbf{LSB}} & \multicolumn{2}{c}{\textbf{PRE}} & \multicolumn{2}{c}{\textbf{RST}} & \multicolumn{2}{c}{\textbf{GST}} \\ \cline{3-16} 
 &  & \textbf{BST} & \textbf{AST} & \textbf{BST} & \textbf{AST} & \textbf{BST} & \textbf{AST} & \textbf{BST} & \textbf{AST} & \textbf{BST} & \textbf{AST} & \textbf{BST} & \textbf{AST} & \textbf{BST} & \textbf{AST} \\ \hline
\textbf{1} & \textbf{Eclipse} & \textbf{0.8187} & 0.8139 & 0.8224 & \textbf{0.8259} & 0.5118 & \textbf{0.5297} & 0.1607 & \textbf{0.1427} & 0.4882 & \textbf{0.4703} & 701.98 & \textbf{676.23} & 231.03 & \textbf{205.28} \\ \hline
\textbf{2} & \textbf{Equinox} & \textbf{0.9554} & 0.9011 & \textbf{0.9356} & 0.9208 & \textbf{0.7149} & 0.6855 & \textbf{0.0882} & 0.1176 & \textbf{0.2851} & 0.3145 & \textbf{37.76} & 41.65 & \textbf{11.68} & 15.57 \\ \hline
\textbf{3} & \textbf{Mylyn} & \textbf{0.8493} & 0.8165 & \textbf{0.8373} & 0.8233 & \textbf{0.6789} & 0.6424 & \textbf{0.105} & 0.1415 & \textbf{0.3211} & 0.3576 & \textbf{467.42} & 520.57 & \textbf{152.85} & 206 \\ \hline
\textbf{4} & \textbf{PDE} & 0.8486 & \textbf{0.9021} & 0.8815 & \textbf{0.902} & 0.7146 & \textbf{0.7758} & 0.1839 & \textbf{0.1227} & 0.2854 & \textbf{0.2242} & 346.27 & \textbf{272.05} & 223.08 & \textbf{148.86} \\ \hline
\multicolumn{1}{l}{} & \textbf{Average} & \textbf{0.868} & 0.8584 & \textbf{0.8692} & 0.8680 & 0.6551 & \textbf{0.6584} & 0.1345 & \textbf{0.1311} & 0.3450 & \textbf{0.3416} & 388.36 & \textbf{377.63} & 154.66 & \textbf{143.93} \\ \hline
\end{tabular}
}
\end{table*}
\begin{figure*}[ht]
\begin{subfigure}{.5\textwidth}
  \centering
  \includegraphics[width=\linewidth]{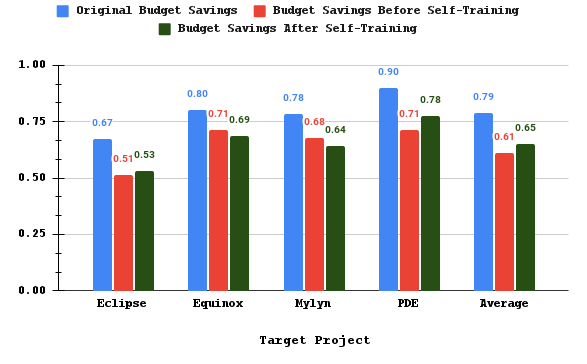}
  \subcaption{The original and predicted percentages of the Budget Savings measure}
  \label{BudgetSavings}
\end{subfigure}%
\begin{subfigure}{.5\textwidth}
  \centering
  \includegraphics[width=\linewidth]{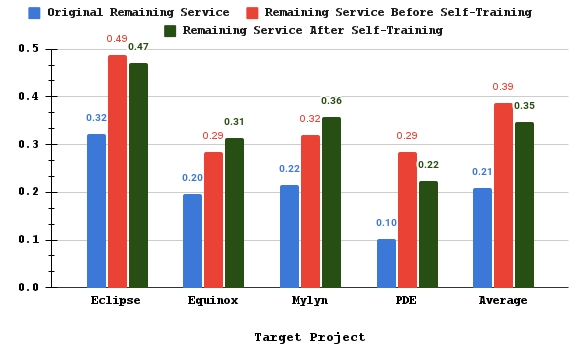}
  \subcaption{The original and predicted percentages of the Remaining Service measure}
  \label{RemainingService}
\end{subfigure}
\caption{Average saved budget and remaining service performances on all the datasets}
\label{SBAndRS}
\end{figure*}
The table \ref{Performances} presents the performances obtained before and after implementing the decision tree based self-training semi-supervised model. The left column under each measure represents the performance of the decision tree model, which is trained on the original set of labelled data, whereas the right column under each measure represents the performance of the proposed model on the combined pseudo-labelled data.

In this empirical study, contradictory predictions are observed between the traditional and proposed measures. It is observed from the table \ref{Performances} that, on an average, the values of the traditional measures such as \textit{Accuracy} and \textit{F-Measure} are reduced after implementing the proposed self-training approach. Before implementing the self-training model, the average \textit{Accuracy} is observed as 0.868. It is reduced by 0.0096 points after implementing the self-training model. Similarly, before implementing the self-training model, the average \textit{F-measure} was observed as 0.8692. It is reduced by 0.0012 points after implementing the self-training model. 

On the contrary, the proposed measures show the real-benefits of the self-training model. From the table \ref{Performances} it is observed that, before implementing the self-training model, the average \textit{PSB} is observed as 0.6551. It improved by 0.0033 points after implementing the self-training model. Hence, the testers need not visit nearly 65.84\% of the total code to discover the defects. Similarly, the self-training model tries to reduce the loss in the saved budget measure on the target projects. From the table \ref{Performances} it is observed that, before implementing the self-training model, the average \textit{LSB} is observed as 0.1345. It is reduced by 0.0034 points after implementing the self-training model. That is, the testers still need to investigate nearly 13.11\% of the total code for the non-existing defects. Since, \textit{LSB} imposes unnecessary code reviews, minimising this value is the major objective of the prediction model.

Similarly, the measures such as \textit{PRE}, and \textit{RST} also show the real-benefits from the self-training model. To estimate the value of \textit{RST}, we set $\delta$ to 100 to make implementation easier. From the table \ref{Performances} it is observed that, before implementing the self-training model, the average \textit{PRE} is observed as 0.3450. It is reduced by 0.0034 points after implementing the self-training model. Therefore, the testers need to spend nearly 34.16\% on the total code to discover the total defects in all the target projects. Here, for the remaining 34.16\% of the total code, the testers need to spend nearly, 377.63\% average project hours to identify/remove the total defects in the system. Before implementing the self-training model, the testers would have to spend 388.36\% average project hours to identify/remove the total defects in the system.

However, the measures \textit{PRE} and \textit{RST} also include false positives (\textit{FP7-FP10}). In this regard, the measure \textit{GST} computes the unnecessary project hours that the tester is spending on the false positives (\textit{FP7-FP10}) modules. From the table \ref{Performances} it is observed that, before implementing the self-training model, the average \textit{GST} is observed as 154.66 project hours. It is reduced by 10.73 hours after implementing the self-training model. Since these false positives degrade the value of the predictions, minimising the modules predicted into this category is also the major objective of the prediction model.

On the other hand, the figures \ref{BudgetSavings} and \ref{RemainingService} show how far the machine learning model achieved the respective performances in terms of the percent of original budget savings and original service time. The figure \ref{BudgetSavings} presents the percentages of original budget savings, budget savings before the self-training, and budget savings after the self-training. The figure \ref{RemainingService} presents the percentages of original remaining service, remaining service before the self-training, and remaining service after the self-training. From the figure \ref{BudgetSavings}, it is observed that, on an average, even though the self-training model was able to save 65\% of the total budget, there is still a gap of 14\% from the original budget savings. Similarly, from the figure \ref{RemainingService}, it is observed that, on an average, even though the self-training model imposes 35\% of the service time on the tester, it is still possible to reduce another 14\% of the code coverage through the predictions.
\subsection{Discussion}
\label{Discussion}
\subsubsection{On the Use of Self-Training Semi-Supervised Learning Model}
% Please add the following required packages to your document preamble:
% \usepackage{multirow}
\begin{table}[ht]
\centering
\caption{Supplementary results: Saved budget and remaining edits}
\label{SBRE}
\resizebox{\linewidth}{!}{
\begin{tabular}{crrrrr}
\hline
\multirow{2}{*}{\textbf{Target Project}} & \multicolumn{1}{c}{\multirow{2}{*}{\textbf{Total LoC}}} & \multicolumn{2}{c}{\textbf{Saved Budget}} & \multicolumn{2}{c}{\textbf{Remaining Edits}} \\ \cline{3-6} 
 & \multicolumn{1}{c}{} & \multicolumn{1}{c}{\textbf{BST}} & \multicolumn{1}{c}{\textbf{AST}} & \multicolumn{1}{c}{\textbf{BST}} & \multicolumn{1}{c}{\textbf{AST}} \\ \hline
\textbf{Eclipse} & 143788 & 73590 & \textbf{76165} & 70198 & \textbf{67623} \\ \hline
\textbf{Equinox} & 13245 & 9469 & 9080 & 3776 & 4165 \\ \hline
\textbf{Mylyn} & 145589 & 98838 & 93532 & 46742 & 52057 \\ \hline
\textbf{PDE} & 121333 & 86706 & \textbf{94128} & 34627 & \textbf{27205} \\ \hline
\textbf{Total} & 423955 & 268603 & \textbf{272905} & 155343 & \textbf{151050} \\ \hline
\end{tabular}
}
\end{table}
The works \cite{tanha2017semi, zou2018unsupervised, zou2019confidence, yu2020fine, oymak2020statistical} show the advantages of using the self-training model. As an example, from the table \ref{Performances}, it is observed that, the proposed decision tree based self-training semi-supervised learning model accounts for an improvement in the performance on the majority of the target projects.

Precisely, the experimental results show that, on an average, the self-training model with decision tree learning is showing its improvement in saving the project budget of 1.6016\% cost units (4302 cost units) and reducing the remaining service time of 2.8421\% of hours. That is, from the table \ref{SBRE}, it is observed that, on an average, self-training saves 272,905 cost units. On the original labelled set, the decision tree model tries to save only 268,603 cost units out of 423,955 cost units. Similarly, the self-training model decreases the remaining edits to 151,050 time units from 155,343 time units. Hence, with the inclusion of more unlabelled data, there is a possibility to improve the final prediction performance.
\subsubsection{On the Project Specific Performance Measures}
As discussed in section \ref{evalMeasures}, the performance measures are the key elements to describing the success of the prediction model. The traditional measures (such as \textit{Precision, Recall, F-Measure} etc.) are commonly used in most machine learning applications. However, only a few measures are appropriate for the working application. In this regard, the proposed measures (given in section \ref{evalMeasures}) will provide relevant information to the project managers from the obtained predictions.

For instance, to understand the real benefits of the prediction model, let us assume the model is implemented after self-training. From the table \ref{Performances}, it is observed that the self-training model achieved a total budget savings (PSB) of 65.84\%, requiring the test team to spend the quality amount of time (PRE) only on the remaining 34.16\% LoC. For this, the testers have to spend 377.63 project hours (RST) searching for the defects in the respective modules. However, the value of PRE also contains information about the false positives (of 13.11\% LSB). Therefore, out of 377.63 project hours, the testers are allocated unnecessarily to the software modules for nearly 143.93 extra project hours (GST).

In summary, as described above, the proposed measures show an ability to interpret the results in terms of the project-specific attributes. Hence, in addition to the traditional measures, we recommend using the proposed measures to present the prediction performances.
\section{Threats to Validity}
\label{threats}
This section presents the threats that may challenge the validity of the proposed approach.
\subsection{Construct Validity}
\label{Construct Validity}
On the basis of the acceptance threshold $\Gamma$, the self-trained semi-supervised model assigns high-confidence class labels to unlabeled instances. In \cite{oymak2020statistical}, Oymak et al. provided an analysis for selecting the proper value of $\Gamma$ to obtain the best predictions on the test set. The experimental evaluations conclude that, upon selecting the higher value of the $\Gamma$ can improve the final performance of the model. Hence, in this work, we evaluated all the projects using the proposed model by setting the value of the acceptance threshold ($\Gamma$) at 0.99. Nevertheless to say, the final performance of the proposed model may differ after experimenting with different values of the acceptance threshold $\Gamma$.
\subsection{Internal Validity}
\label{Internal Validity}
To balance various classes, this work utilises the concept of ADASYN (a variant of SMOTE). Nonetheless, any of the other approaches such as SMOTE, SMOTE-NC, Borderline-SMOTE, etc. are the other alternate choices to observe a change in the prediction performance of the self-training model. Implementing the ADASYN technique is a random choice and was included in the proposed approach based on the ablation study. Apart from implementing the oversampling techniques, the performance of the proposed model may also be affected by the use of undersampling techniques.

In self-training, the pseudo-labels are assigned to the unlabelled instances at each iteration. In the process of generating pseudo-labels, the self-training operates on the basis of classification and regression trees (an approach of decision tree classifier). We suspect that the prediction performance may vary up-on implementing the self-training with the other base-learners, instead of the decision tree classifier.
\subsection{External Validity}
\label{External Validity}
External validity refers to the generalizability of the performance results obtained from the proposed approach. In this work, we have conducted an experiment on the 527 unlabeled instances collected from the four software systems. The above defect count might not be representative enough. Hence, to avoid this threat in the future, we will extend the proposed approach to more unlabelled instances which are collected from the heterogeneous projects.
\section{Conclusion}
\label{conclusion}
This work is targeted to bridge the research gap by providing a classification model to find the severity of the defective software modules using a metric-based solution and to provide a project-specific analysis from the predictions of the proposed approach.

For the first objective, we proposed a self-training semi-supervised learning approach based on the decision tree model. The self-training method is an iterative approach in which the base model (decision tree classifier) first builds confidence on unlabeled instances by assigning class labels and then, using the acceptance threshold ($\Gamma$), selects the high-confidence instances from the unlabeled data at each iteration.  The selected pseudo-labelled instances are then appended to the original set of labelled instances, again to build the new classifier. In the end, to know whether the self-training model successfully assigns the labels to the unlabelled instances, the combined pseudo-labelled dataset is then given as input to train decision tree classifier.

To understand how far the prediction results are helpful to the project managers, in this work, we proposed five project specific measures, such as, \textit{the Risk-Factor (RF)}, \textit{the Percent of Saved Budget (PSB), the Loss in the Saved Budget (LSB)}, \textit{the Remaining Service Time (RST)}, and \textit{Gratuitous Service Time (GST)}. \textit{The Risk-Factor} is calculated to measure the amount of risk that the project encounters if a false negative case occurs. Whereas \textit{The PSB} and \textit{LSB} are indicative of the savings and the loss of the original savings in the project, respectively. \textit{The RST and GST} measure the amount of time still required to service the damaged code and unnecessary time spent on the project, respectively, if the prediction model exhibits any non-true negative instances.

In this article, we recommend using the self-training semi-supervised learning approach to predict the severity of the defective software modules on the large set of unlabeled data since the experimental evaluations (using the inbuilt decision tree model) show the benefits (in terms of all the performance measures) of the proposed approach. Similarly, we also recommend using the proposed performance measures in the future defect severity prediction studies in order to know the clear project benefits from the prediction model.

Due to the rapid growth in the development of software systems in the future, we extend this study to find the severity of the defective software modules in heterogeneous projects.

\bibliographystyle{IEEEtran}
\bibliography{SDSP}
\end{document}